\documentclass{llncs}

\usepackage{graphicx}
\usepackage[T1]{fontenc}
\usepackage{enumerate}
\usepackage[table,xcdraw]{xcolor}
\usepackage{amssymb}
\usepackage{mathrsfs}
\usepackage{amsmath}
\usepackage{enumitem}
\usepackage{color}
\usepackage{minibox}

\begin{document}

\pagestyle{headings} 

\addtocmark{Study on Resource Efficiency of Distributed Graph Processing}
\title{Study on Resource Efficiency of Distributed Graph Processing}
\titlerunning{Study on Distributed Graph Processing} 

\author{Miguel E. Coimbra \and Alexandre P. Francisco \and Lu\'{i}s Veiga}
\authorrunning{Miguel E. Coimbra et al.}
\tocauthor{Miguel E. Coimbra, Alexandre P. Francisco, and Lu\'{i}s Veiga}
\institute{INESC-ID Lisboa / Instituto Superior T\'{e}cnico, University of Lisbon, Portugal\\
\email{miguel.e.coimbra@tecnico.ulisboa.pt \\ \{aplf, luis.veiga\}@inesc-id.pt}
}
\maketitle

\begin{abstract}
Graphs may be used to represent many different problem domains -- a concrete example is that of detecting communities in social networks, which are represented as graphs.
With big data and more sophisticated applications becoming widespread in recent years, graph processing has seen an emergence of requirements pertaining data volume and volatility.
This multidisciplinary study presents a review of relevant distributed graph processing systems.
Herein they are presented in groups defined by common traits (distributed processing paradigm, type of graph operations, among others), with an overview of each system's strengths and weaknesses.
The set of systems is then narrowed down to a set of two, upon which quantitative analysis was performed.
For this quantitative comparison of systems, focus was cast on evaluating the performance of algorithms for the problem of detecting communities.
To help further understand the evaluations performed, a background is provided on graph clustering.

\keywords{distributed graph processing; graph theory}
\end{abstract}

\section{Introduction}\label{sec:introduction}

Graphs are, and have been for many decades, an ubiquitous mental artifice to model a plethora of real-world problems.
An article written by Leonhard Euler on the \textit{Seven Bridges of K\"{o}nigsberg} problem of mathematics over two hundred years ago is regarded in graph theory to be the one that formally pioneered the field (N. C. Jones and P. A. Pevzner~\cite{jones2004introduction}).
For example, a very important algebra operation is that of inverting a matrix: fifty years ago a proposal was made for a method~\cite{nathan1967inversion} which represents a matrix as a graph, in order to rapidly perform this operation.
Graphs' more recent range of usage covers, to name a few examples, analyzing the structure of the World Wide Web~\cite{Boldi:2004:WFI:988672.988752}, bio-informatics data representation via \textit{de Bruijn} graphs~\cite{ACombinatorialProblem1946} in metagenomics~\cite{Li01022010},~\cite{18349386}, the structure of distributed computation~\cite{Malewicz:2010:PSL:1807167.1807184},~\cite{Han:2014:ECP:2732977.2732980} and data modeling to attempt to \textit{predict the future}, as is the case with applications of Markov Models and Hidden Markov Models (K. Murphy~\cite{Murphy:2012:MLP:2380985}, Ch. 17).
Graph theory is thus, in itself, both a means and an end in many scenarios.\\
\indent This work aims to highlight recent contributions in distributed systems addressing graph processing, targeting community detection as a problem of graph clustering.
Performed experiments were aimed at analyzing the performance of two distributed graph processing systems in light of a label propagation algorithm which allows for community detection.
These two systems chosen for quantitative comparison were the result of a broader qualitative analysis of several graph processing systems.
A background on community detection as a problem of graph clustering and label propagation is provided in Sec.~\ref{sec:sec:domain}, with an overview on four methods.
For this study, emphasis was put on the analysis of label propagation, in light of having properties that make it friendly to parallel approaches.
The study was conducted as part of a larger context of big data, dynamic graphs and the emerging roles of approximate and incremental computing.

\subsection{Motivation}\label{sec:sec:motivation}

There is an interest in developing means of improving the computational speed of recalculating metrics and properties as a consequence of dynamism in graphs.
As graphs represent data ranging from historical sciences to social networks, with varying size, differing needs of result accuracy and speed of recalculation are bound to exist.\\
\indent Given the increasing dimension of data sets represented as graphs in the advent of big data, how far can one relax from exact to approximate computing?
How may we reliably define and ensure controlled error bounds given the volatility of graphs such as those representing social networks?
And to what extent can performance be gained with a trade-off in accuracy in distributed system graph-processing scenarios?
Employing distributed systems, how may one scale-out methods of graph clustering (potentially in line with statistical and learning techniques) to process large graphs while harnessing the computational power of distributed computing infrastructures?
How easy is it to adopt incremental solutions for the processing of dynamic graphs?
Could we update existing graph processing results (or graph clusters, in the scope of this document) based solely on previous results and a given change, or set of changes, in the graph?
Establishing a basis for solving these challenges would pave the way for greater performance and resource-efficiency in the analysis of many graph-based big data scenarios.
This study is therefore aimed at gaining knowledge of the most suitable platforms to study these questions.

\subsection{Challenges}\label{sec:sec:challenges}

These questions, among others, form a body of non-trivial properties which must be accounted for in distributed graph processing.
The problem of non-overlapping graph clustering is no exception to these questions.
Studying the structure of different \textit{hard} clustering algorithms to harness the power of distributed systems is a preliminary step in that direction.
Effectively, in a wider body of work, the case of graph clustering was chosen as a basis to perform studies in line with the described motivations.
A detailed study on distributed large graph processing is something that requires establishing a basis to develop further work.\\
\indent There exists a plethora of commercial and open-source distributed systems (some are general-purpose with graph processing modules, others are dedicated to graph processing) which may be used as a platform for conducting further studies.
This work used graph clustering as an initial case study for graph processing (in the context of large graph distributed processing), with the aim of responding to two challenges: \textit{1)} assessing the most desirable system(s) for graph processing to use for further work; \textit{2)} attempting to gain knowledge of the graph clustering case to contemplate other graph problem classes in the literature (under the advent of incremental, approximate and streaming computing), such as incremental local metric calculation (e.g., incremental triangle counting), ranking (e.g., how does a given metric change in response to an operation triggering a propagation in the graph) and global metric calculation (e.g., those which are derived from shortest-path calculations, which are challenging for static graphs -- considering dynamic graphs would further complicate things).

\section{Background and Related Work}\label{sec:sec:domain}

Graph clustering is a problem which arises in several domains, with the goal of finding groups that are homogeneous (in the sense that the vertices contained in a graph cluster probably share common properties) and under certain separation criteria.
Algorithms for this may try to optimize for a specific set of parameters or try to apply knowledge about the underlying data.
Clustering may vary depending on whether \textit{soft} or \textit{hard} clustering is considered.
This is usually formalized by defining a quality measure which may differ with the problem domain.
When analyzing graph community structure, an attempt is made to evaluate the separation of sparsely connected dense sub-graphs from each other.
Formally, let $G = (V, E) $ be an undirected graph.
\label{def:clustering}
A non-overlapping clustering or partition $P$ of $G$ is a collection of sets \{$V_1,...,V_k$\}, with $k \in \mathbb{N}$, such that $V_i \neq \emptyset$ for $1 \leq i \leq k, V_i \cap V_j = \emptyset$ for $1 \leq i < j \leq k$, and $\bigcup_{1 \leq i \leq k}V_i = V$.\\
\indent Ideally, a system would be able to compute the changes impacting the graph with \textit{enough speed}, that is, so that the human end-users of the system do not suffer a negative experience (typically manifested by their awareness of the computational delay).
Likewise, the clients of these computations (performed in \textit{Software-as-a-Service} infrastructures for example) could be automated software components which must have their data queries solved under strict time constraints.
Furthermore, if the graph represents a social network, such as Facebook with its vast user base and colossal amount of edges and graph properties~\cite{Ching:2015:OTE:2824032.2824077}, it is highly unrealistic to assume that graph operations occur sequentially, which means that an operation will have consequences in terms of both the amount of data affected and the way different agents interact in the graph.
The degree to which hardware resources can be employed for parallel processing in graph clustering algorithms depends heavily on the nature of the computation performed.
This section presents some of the most relevant graph clustering methods in the literature (as far as the author knows):
\begin{itemize}[leftmargin=*]
\item \textbf{Louvain Method}. 
A greedy optimization method, able to identify communities (again, a subset of clustering variants) in large networks.
For an undirected graph $G = (V, E)$, this technique  has an apparent\footnote{No formal proof exists in the literature, this is based on experimental results by researchers.} computational complexity of $O(|V|\ log\ |V|)$; it requires effort \textit{almost} linear in the $|V|$ number of vertices of the graph.
It is not unusual to encounter the term \textit{Louvain Modularity} in the literature.
That is the name of the metric which the Louvain Method attempts to optimize~\cite{Blondel08fastunfolding}.
It has been used to explore the problem of partitioning social networks onto different machines and to identify dynamic communities in dynamic social networks (inherent in mobile networks).
The modularity of a partition is defined as a scalar value in the interval $[-1, 1]$, measuring the amount of links contained inside communities versus the links across communities.
Modularity in fact measures the fraction of edges in the network that connect within-community edges minus the expected value of the same quantity in a graph network with the same community divisions, but random vertex connections.
If the number of edges connecting same-community vertices is much lower than that of the random network, modularity can become negative.
However, most methods that calculate modularity typically do not produce values lower than those obtained when each vertex is its own community (and that scenario would yield a negative value).
Values are usually between the value of that scenario and the upper bound of 1.
In the literature one may find a definition of modularity such as the following:
\begin{equation}
	Q = \dfrac{1}{2m}\sum_{i, j}\big[A_{ij} - \dfrac{k_{i}k_{j}}{2m} \big]\delta(c_{i} ,c_{j})
\end{equation}
In it, $m$ and $k_{i}$ are defined as:
\begin{equation}
	m = \dfrac{1}{2}\sum_{ij}A_{ij}\qquad	k_{i} = \sum_{j}A_{ij}
\end{equation}
Where \textbf{A} is a weight matrix with element $A_{ij}$ representing the weight of an edge between a vertex $i$ and a vertex $j$, $m$ is half of the sum of all elements of $A$ and $k_{i}$ is the sum of the weights of the edges attached to vertex $i$ (the sum of row $i$ of matrix \textbf{A}). $\delta$ is the Dirac delta function.
Modularity $Q$ has been used as both a quality assessment of partitioning methods and also as an optimization problem in itself.
The method works by iterating until no further increases in modularity occur.
Initially, each vertex is its own community.
For every vertex $i$, the potential gain in the modularity from moving a vertex out of its own community (thus acting as an isolated vertex) and into each of its neighbors' communities is calculated through a similar formula.
In the literature, this main method of operation is said to be using a heuristic called \textit{local moving heuristic}~\cite{Waltman2013}.
Essentially, if vertex $i$ is moved into\footnote{The removal of a vertex $i$ from a community $C$ also has a similar formula.} a community $C$, the change in modularity may be obtained with the following expression:
\begin{equation}
\Delta\ Q = \bigg[\dfrac{\sum_{in} + k_{i, in}}{2m} - \bigg(\dfrac{\sum_{tot}+k_{i}}{2m}\bigg)^2\bigg] - \bigg[\dfrac{\sum_{in}}{2m} - \bigg(\dfrac{\sum_{tot}}{2m}\bigg)^2 - \bigg(\dfrac{k_{i}}{2m}\bigg)^2\bigg]
\end{equation}
For this expression, $\sum_{in}$ is the sum of the weights of the edges in community $C$ and $\sum_{tot}$ is the sum of the weights of the edges incident to vertices in $C$ and $k_{i,in}$ is the sum of the weights of edges between vertex $i$ and vertices of community $C$.
Other symbols retain the previous meaning.
Phase two consists in building a new network where nodes are new communities found during the previous phase.
Weights of links between new nodes (which are the communities established during phase one) are the sum of weight of links between nodes in the two corresponding communities.
\begin{figure}[t]
\quad\quad\quad\quad$G_{0}:Q_{0}${\hspace{.24\linewidth}}
$G_{1}:Q_{1}${\hspace{.25\linewidth}}
$G_{2}:Q_{2}${\hspace{.33\linewidth}}\\
\includegraphics[width=\textwidth]{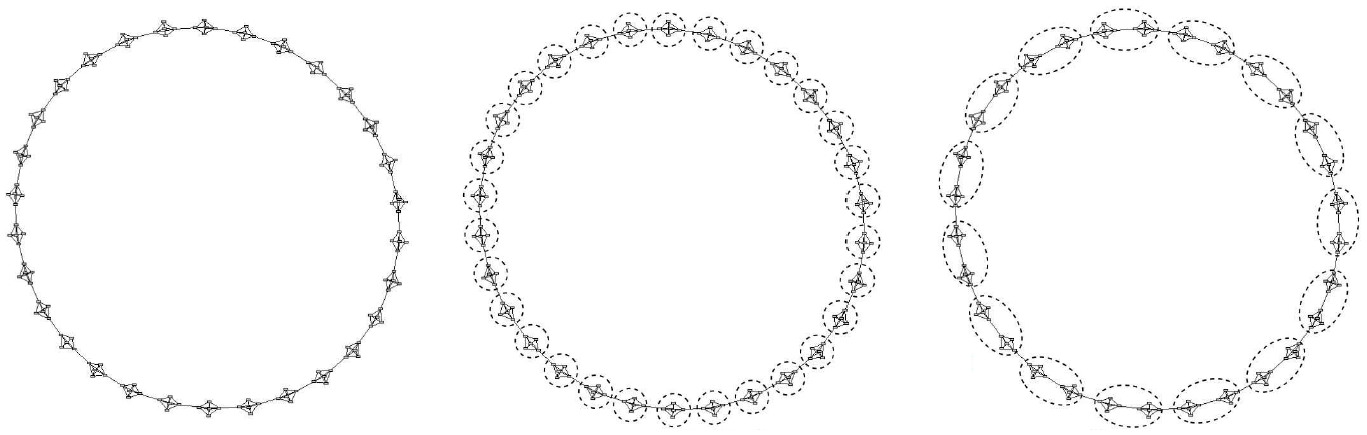}
\caption{
\label{fig:louvain}
Repetition of the Louvain Method for the 30-clique example shown in~\cite{Blondel08fastunfolding}.
After the first passage of the two phases of the algorithm, there is graph $G_{1}$ with thirty communities.
Executing one more phase leads to even \textit{bigger vertices} in graph $G_{2}$.
Visually, iteration of the method occurs from left to right, with ever-increasing modularity: $Q_{0}<Q_{1}<Q_{2}$.}
\centering
\end{figure}
As shown in Fig.~\ref{fig:louvain}, a visual notion of hierarchy emerges as the algorithm progresses.
This algorithm is unsupervised and it is considered fast; albeit proof of linearity is missing, simulations on large ad-hoc modular networks suggest its complexity is linear on typical and sparse data~\cite{Blondel08fastunfolding}.\\
\indent The aforementioned method, as a modularity optimization technique, incurs resolution limitations regarding the size of communities, as noted in~\cite{Fortunato:2007p5549}.
However, it is important to retain that this method yields different community resolution levels as iterations progress.
As such, and as its authors suggest, the sequence of iterations produces heterogeneous community configurations, depending on the execution step.
This could allow end-users to \textit{zoom in} the network and observe it with a specific resolution, depending on how much the two phases are iterated.
\item \textbf{Label Propagation}. Label propagation algorithms do not on their own involve modularity optimization.
Their application for community detection is therefore scale-independent and not affected by the previously mentioned community size resolution limit.
Label propagation typically starts with each vertex being associated to a unique label.
In every iteration, the label of each vertex is updated by choosing the label that most of its neighbors have (maximal label).
Tie-breaking between multiple maximal labels can be done with random label picking.
An example of a basic label propagation algorithm proposal can be found in~\cite{PhysRevE.79.066107}.\\
\indent In more elaborate variants of label propagation, score measures may be employed to account for the weights of the edges~\cite{PhysRevE.76.036106}.
Scoring was incorporated to deal with \textit{epidemic labels} (covering for example over 50\% of the vertex count), which override smaller communities lacking strong-enough links.\\
\indent A variant of label propagation is the Layered Label Propagation (LLP) \break method~\cite{Boldi:2011:LLP:1963405.1963488}, devised to tackle the problem of resolution limits.
It verified the hypothesis posed by the authors of the Louvain Method -- they suggested the generation of different community resolution hierarchy levels, depending on the execution step.
LLP is a parameter-free scalable clustering algorithm that can reorder large graphs, with vertex count $|V|$ in the order of billions.
Originally, it was devised whilst exploring ways to improve graph compression (this work was a ramification of prior research on web graph compression) by harnessing graph ordering.
A partition induces an ordering, and the appropriateness of the partition may be measured by comparing an original order with the possible refinements brought about by the induced ordering.
For a given graph $G = (V, E)$ and a partitioning $\mathscr{U}$, the entropy as a function of the partition (please refer to \cite{Boldi:2011:LLP:1963405.1963488} for additional in-depth detail on mathematical clarity) is defined as:
\begin{equation}
\mathbb{H}(\mathscr{U}) = -\sum_{i=0}^{R}\dfrac{|\mathscr{U}_{i}|}{|V|}log\bigg(\dfrac{|\mathscr{U}_{i}|}{|V|}\bigg)
\end{equation}
Each sub-component of the partition is identified by $\mathscr{U}_{i}$ and $R$ is the number of sub-components.
The contextual basis of Layered Label Propagation was the web graph, in which vertices represent web pages (URLs).
Considering that, it was found in previous work \cite{Boldi03thewebgraph} that lexicographical ordering of vertices (based on the URLs) yielded better compression capabilities.
In that regard, and to assess the power of Layered Label Propagation (and other methods which produce orderings), the authors make use of another statistical artifice, the variation of information between a given intrinsic partition $\mathscr{H}$ and a respective refinement based on an ordering $\pi$ (which translates into a partition represented as $\mathscr{H}_{|\pi}$).
After some manipulations, they arrive at the following formula:
\begin{equation}
\mathbb{VI}(\mathscr{H}, \mathscr{H}_{|\pi}) = \mathbb{H}(\mathscr{H}_{|\pi}) - \mathbb{H}(\mathscr{H})
\end{equation}

The core principle of this heuristic is that a lower variation of information implies that the ordering is more \textit{desirable} or \textit{of better quality}.
The major fact pertaining this method is that initially it was motivated by the study of web graphs, but it was also applied to social network data sets.
Making use of the previous formulations, and based on generic label propagation algorithms in the literature, the Layered Label Propagation algorithm was implemented.
It is an iterative algorithm that produces a sequence of node orderings.
For each iteration, the Absolute Potts Model (APM) algorithm~\cite{PhysRevE.81.046114} is executed.
APM has a resolution parameter $\gamma$ which describes a specific community resolution of the graph.
Intuitively, as observed by the authors, optimality as a notion does not apply to $\gamma$: different values for it simply produce descriptions of the graph at various resolution levels.
Lower values of $\gamma$ showcase a coarser graph structure with sparser and bigger clusters, and increasing the value produces smaller and denser clusters, uncovering a finer structure.
The algorithm is such that it produces an order of the graph keeping nodes with the same label close to one another in the ordering.
For $K$ iterations, each iteration $i$ will be subjected to a chosen $\gamma_{i}$.
In line with the previous note on the (lackluster) merit of optimizing $\gamma$, rather than trying to find an optimal value for each iteration, $\gamma_{i}$ is instead uniformly drafted at random from the set $\{0\}\cup\{2^{-i}, i = 0, ... , K\}$.
This conclusion was reached through experimental results; optimizing $\gamma$ on each iteration was not computationally efficient.
\item \textbf{$K$-means clustering}.
Another variant of clustering, rooted in signal processing, which has the objective of partitioning $n$ observations into $k$ clusters, such that each observation belongs to the graph cluster with the nearest mean.
The $k$-means is a point-assignment type of family of clustering algorithms (a heuristic-based variant is Lloyd's algorithm).
It works under the assumption that the number of clusters $k$ is known $a\ priori$ and execution occurs in the context of an Euclidean space.
Despite this, methods exist which entail a trial-and-error approach to determining $k$.
A high-level conceptual description of the algorithm is presented.

\noindent\makebox[\linewidth]{\rule{\textwidth}{0.4pt}}
\textbf{Basic $k$-means pseudo-code}\\
\noindent \textbf{INPUT:}\quad\ \ \ Graph $G = (V, E)$\\
\noindent \textbf{OUTPUT:}\ \ Collection of sets $\{P_{1}, P_{2}, ..., P_{k}\} : P = P_{1} \cup P_{2} \cup ...\cup P_{k} = V$, \textbf{as per Sec.~\ref{def:clustering}}\\
\noindent1: $L\quad\longleftarrow\quad$CHOOSE($V, k$)\\
\noindent2: $P\quad\longleftarrow\quad$DEFINE-CLUSTERS($L, V$)\\
\noindent3: for each point $w \in V$ \textbackslash\ $L$:\\
4:\qquad $index\quad\longleftarrow\quad$\ FIND-CLOSEST-CENTROID($P, L, w$)\\
5:\qquad $P_{index}\quad\longleftarrow\quad$$P_{index} \cup\ w$\qquad\qquad\qquad\qquad\qquad\qquad\qquad\qquad\ \textit{(Expectation)}\\
6:\qquad UPDATE-CENTROID($P_{index}$)\qquad\qquad\qquad\qquad\qquad\qquad\qquad\ \textit{(Maximization)}

\noindent\makebox[\linewidth]{\rule{\textwidth}{0.4pt}}

\noindent It is suggested (Chapter 7 of~\cite{leskovec2014mining}) that $k$-means should be executed with increasing values of $k = 1, 2, 4, 8, ...$.
The formulation behind the suggestion is that eventually, a value $v$ will be found such that, for variations of $k \in [v, 2v]$, there will be little change in the measure of graph cluster cohesion used.
Lines five and six of the presented pseudo-code should receive particular attention: in reality, the $k$-means algorithm iteration cycle can be described as having an expectation step (line five) followed by a maximization step (line six).
This constitutes an incarnation of the expectation-maximization algorithm (EM for short).
In~\cite{Murphy:2012:MLP:2380985}, the expectation-maximization introduction defines the following equation, which is the expected complete data log likelihood:
\begin{equation}
Q(\theta, \theta^{t-1}) = \mathbb{E}[l_{c}(\theta)|\mathcal{D},\theta^{t - 1}]
\end{equation}
In this equation, $\theta$ represents the unknown state of nature, $\mathcal{D}$ the observed data, $t$ is the current iteration number and $Q$ is called the auxiliary function\footnote{Concretely, what is calculated in the expectation step are the fixed data-dependent parameters of the function $Q$.} .
It makes use of the complete data log likelihood $l_{c}$, defined as:
\begin{equation}
	l_{c}(\theta) \triangleq \sum_{i = 1}^{N}\log p (x_{i}, z_{i}|\theta)
\end{equation}
The missing data is $z_{i}$, in this case representing the assignment of vertices to communities.
Here $z_{i}$ represents \{$q$\}, the model's parameters \{$\pi$\} are implicit and there is a direct mapping to the amount of clusters \{$k$\}.
$K$-nearest neighbors is a supervised learning method for classification.

\item \textbf{Nearest neighbors clustering}. Delving further into clustering, it may be interpreted as a general baseline algorithm for minimizing arbitrary clustering objective functions.
It is in fact a single-link clustering technique: an agglomerative (bottom-up hierarchical clustering) method.
For each step, the closest pair of clusters (pair of elements in the data set in the first iteration) are chosen, with the clusters being merged.
Naive versions of this algorithm have a computational complexity ranging from $O(|V|^2)$ to $O(|V|^3)$ which, coupled with its greedy (without making use of backtracking mechanisms) nature, constitute relative drawbacks.
The algorithm begins with every vertex $v$ as its own graph cluster.
After $|V| - 1$ iterations have ensued, a single graph cluster (binary tree) remains.
It should then be \textit{cut} at a parametrized depth to produce a specific partitioning (the result of clustering may be seen using a representation such as a dendrogram).
A visual example of such a distance-specific hierarchy is displayed in Fig.~\ref{fig:dendrogram}.
A major configuration of this algorithm resides in the choice of distance function between clusters~\footnote{Variants exist depending on the definition of graph cluster distance: single-link or minimum distance, complete-link or maximum distance, average distance and mean distance.}.
This graph cluster distance may be mathematically written as:
\begin{equation}
D(A, B) = \underset{a \in A, b \in B}\min\ d(a, b)
\end{equation}
Where $A$ and $B$ are any two clusters and $d(a, b)$ represents the distance between two elements $a$ and $b$ of each graph cluster.
The description provided so far is a naive version, sharing many similarities with Kruskal's~\cite{10.2307/2033241} minimum spanning tree algorithm.
Improved methods exist: in particular, it has been pointed out in the literature that better results may be achieved by running Prim's algorithm~\cite{BLTJ:BLTJ1515} for spanning trees before Kruskal's.
The idea is to use Prim's algorithm without priority queues, using $O(|V|)$ space and $O(|V|^2)$ execution time.
This results in a minimum spanning tree of the elements and distances (this tree is actually a sparse graph).
Afterward, Kruskal's algorithm may be put to use over the edges of the minimum spanning tree, finally producing the clustering; additional space requirements of $O(|V|)$ and $O(|V|\log|V|)$ ensue.
Ultimately, this variant has a temporal complexity improvement, ensuring its bound is $O(|V|^2)$.
\begin{figure}[t]
	\centering
	\includegraphics[width=0.5\textwidth]{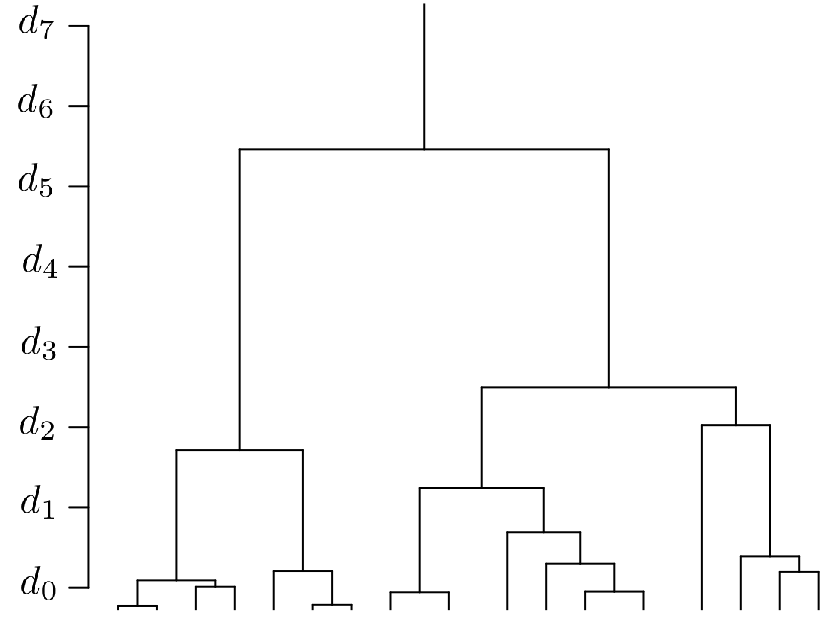}
	\caption{A dendrogram distance-based \textit{cut} example which was handed a graph cluster at iteration number $|V| - 1$. As the \textit{cutting} distance $d$ is changed, one operates with different degrees of clustering resolution.}
	\label{fig:dendrogram}
\end{figure}
\end{itemize}

\section{Relevant Systems Analysis}\label{sec:related}

The last decade has seen a boom in graph processing technologies.
Different application domains have heterogeneous needs.
On the one hand, the focus may be cast on transaction processing and persistence of information.
An example of this is On-Line Transaction Processing (OLTP), which consists of fast query processing (typically measured in transactions per second) while maintaining data integrity. 
OLTP systems are used to record transactional information in real time, and are expected to return responses quickly, usually in milliseconds.
On the other hand, emphasis may be placed on the analytical aspect, which is the case with On-Line Analytical Processing (OLAP).
OLAP systems serve the purpose of answering analytical queries about stored data.
In the case of LinkedIn, analyzing the graph's connectedness and producing reports on metrics like the number of average connections per user is the sort of task that falls in the scope of an OLAP system~\cite{dimiduk2012hbase}.
This type of system is designed to be capable of answering complex queries which involve aggregations.
In this sense, an effective metric is response time.\\
\indent The most widely used graph processing systems in the literature are described in Sec.~\ref{sec:sec:all_systems}. 
Many of these products make use of distributed databases, some of them even supporting different back-end solutions.
This is followed by a description of some of their more relevant storage backends in Sec.~\ref{sec:sec:backends}.
To the best of the author's knowledge, these are the most relevant distributed systems and tools for graph processing.

\subsection{Systems for Graph Processing}\label{sec:sec:all_systems}

Pertaining the most relevant tools themselves, an exhaustive (to the best of the author's knowledge) list is herein presented.
The first items in the list are a group of projects that have been taken under the umbrella of the Apache Foundation in the more recent years.
An attempt was made to produce a list that is relevant not only in terms of impact and widespread usage but also modernity.
These projects are either aimed specifically at graph processing, or are general-purpose computing frameworks which have been used to build such capabilities.\\
\indent Before going into the tools themselves, it is important to mention \textbf{Apache Spark}.
It is a general-purpose open-source cluster computing engine~\cite{Zaharia:2010:SCC:1863103.1863113}, available under the Apache License 2.0.~\footnote{Access date 21 Oct, '16:~\url{http://spark.apache.org/}}
Apache Spark is a computational platform which supports machine learning and has been used for graph processing.
Its concept stemmed from the limitations of Hadoop when contrasted with organizations' data needs.
Effectively, users were finding the need to run, among others, the type of iterative algorithm that exists in machine learning and graph processing.
Sharing data between MapReduce steps is only possible by writing to a distributed file system (which has its own overhead -- taking up more than 90\% running time of common machine learning algorithms~\cite{Zaharia:2010:SCC:1863103.1863113}).\\
\indent Spark dealt with this issue by introducing the concept of Resilient Distributed Datasets (RDDs), a new storage primitive.
RDDs are fault-tolerant collections of objects partitioned across cluster computing nodes, with the ability to be acted upon in a parallel fashion.
Spark allows for maintaining \textit{hot data} in select RDDs in memory, which allows rapid reuse and thus tackles one of the major limitations found in Hadoop.
An important concept associated to RDDs is the \textit{lineage graph}: a tracked graph of the transformations that applied to build the RDD.
With this, when partitions are lost, this lineage graph is used to perform reconstructions from base data.
Some of the concepts underlying Apache Spark were relevant for other solutions which are described in this section.\\
\indent The following three projects are different idea implementations, often with some overlap, under the umbrella of the Apache Foundation.
\begin{itemize}[leftmargin=*]

\item \textbf{Apache Giraph}. An open-source implementation of Google Pregel~\cite{Malewicz:2010:PSL:1807167.1807184},\break tailor-made for graph algorithms.
It was created as an efficient and scalable fault-tolerant implementation on clusters with thousands of commodity hardware, hiding implementation details underneath abstractions.
Work has been done to extend Giraph from the think-like-a-vertex model to think-like-a-graph~\cite{Tian:2013:TLV:2732232.2732238}.
It uses Apache Hadoop's MapReduce implementation to process graphs.
It was inspired by the Bulk Synchronous Parallel (BSP) model.
Apache Giraph allows for master computation, sharded aggregators, has edge-oriented input, and also uses out-of-core computation -- limited partitions in memory.
Partitions are stored in local disks, and for cluster computing settings, the out-of-core partitions are spread out across all disks.\\
\indent Giraph's authors use the concept of \textit{superstep}, which are sequences of iterations for graph processing.
In a superstep $S$, a user-supplied function is executed for each vertex $v$ (this can be done in parallel) that has a status of active.
When $S$ terminates, all vertices may send messages which can be processed by user-defined functions at step $S+1$.
Giraph attempts to keep vertices and edges in memory and uses only the network for the transfer of messages.

\item \textbf{Apache Flink}.
A tool which supports built-in iterations and delta iterations to efficiently aid in graph processing and machine learning algorithms.
It has a graph processing API called Gelly, which comes packaged with algorithms such as PageRank, Single-Source Shortest Paths and Community Detection, among others.
Fault-tolerance mechanisms are present (lightweight snapshots), and they contemplate both streaming and batch processing strategies.
Flink was built with the aim of supporting all data types and providing seamless code integration.
It supports all Hadoop file systems as well as Amazon S3, among others.
Delta iterations are also possible on Apache Flink, which is quite relevant as they take advantage of computational dependencies to improve performance.
It also has flexible windowing mechanisms to operate on incoming data (the windowing mechanism can also be based on user-specific logic).
Researchers are currently looking into extending its \texttt{DataStream} constructs and its streaming engine to deal with applications where the incoming flow of data is graph-based.

\end{itemize}

The following pair of systems is presented together as they are both based on the concept of edge-centric streaming.
The first one introduced the concept (as far as the author knows), while the later harnessed its features while maximizing performance with respect to the usage of storage devices.

\begin{itemize}[leftmargin=*]

\item \textbf{X-Stream}.
Presented in \textit{SOSP 2013}~\cite{Roy:2013:XEG:2517349.2522740}.
X-Stream introduced the concept of edge-centric graph processing via streaming partitions.
Most of its introductory cases were based on a single-machine shared-memory setup.
However, work has been done to test the feasibility of extending its concepts to take advantage of network-based computing resources~\cite{Malicevic:2014:SGP:2592784.2592789}.
X-Stream exposes the scatter-gather programming model and was motivated by the lack of access locality when traversing edges, which makes it difficult to obtain good performance.
State is maintained in vertices.
This tool uses the streaming partition, which works well with RAM and secondary (SSD and Magnetic Disk) storage types.
It does not provide any way by which to iterate over the edges or updates of a vertex.
A sequential access to vertices leads to random access of edges which decreases performance.
X-Stream is innovative in the sense that it enforces sequential processing of edges (edge-centric) in order to improve performance.
It was evaluated with graph algorithms such as Strongly Connected Components (SCC), Single-Source Shortest Paths (SSSP), PageRank, among others.

\item \textbf{Chaos}.
Presented in \textit{SOSP 2015}~\cite{Roy:2015:CSG:2815400.2815408}, this work had its foundations on X-Stream.
On top of the secondary storage studies performed in the past, graph processing in Chaos achieves scalability with multiple machines in a cluster computing system.
It is based on different features: load balancing, randomized work stealing, sequential access to storage and an adaptation of X-Stream's streaming partitions to enable parallel execution.
Chaos uses an edge-centric \textit{Gather-Apply-Scatter} model of computing.
It is composed of a storage sub-system and a computation sub-system.
The former exists concretely as a storage engine in each machine.
Its concern is that of providing edges, vertices and updates to the computation sub-system.
Previous work on X-Stream highlighted that the primary resource bottleneck is storage device bandwidth.
In Chaos, the storage and computation engines' communication is designed in a way that storage devices are busy all the time -- thus optimizing for the bandwidth bottleneck.
\end{itemize}

The next two tools, while having their own differences as far as the computational model is concerned, both support the specification of update functions to run as kernels on each node of a graph.

\begin{itemize}[leftmargin=*]

\item \textbf{GraphLab}.
(\textbf{Turi}) Originally published in \textit{VLDB} \textit{2012}~\cite{Low_graphlab:a}, \cite{Low:2012:DGF:2212351.2212354}, it is a multi-core graph processing abstraction.
This tool appeared as a parallel programming abstraction for iterative algorithms in sparse graphs.
It was developed while targeting common patterns in machine learning.
It consists of a data graph, which codifies a problem's sparse computational structure and directly-modifiable program state, and a shared data table to support global shared state.\\
\indent GraphLab was created to deal with the lack of expressiveness in parallel abstractions such as MapReduce as well as redundant work required by lower-level tools like MPI and Pthreads, in machine learning.
Testing was based on designing and implementing parallel versions of algorithms for belief propagation, Gibbs sampling and lasso regression, among others.
The authors developed both global computation capability (synchronization mechanism -- in GraphLab terms) and local (via update functions).
This tool provides an API which the authors describe to be a potential bridge between the machine learning and systems communities.
In order to provide efficient parallelism, GraphLab introduces variations of consistency models: one for vertices, another for edges and another one called full consistency.
This enables different degrees of parallelism.
GraphLab has been evaluated with applications such as collaborative filtering for Netflix movie recommendations (Alternating Least Squares algorithm), Video Co-segmentation (CoSeg) and Named Entity Recognition (NER).

\item \textbf{GraphX}.
Presented in \textit{OSDI 2014}~\cite{Gonzalez:2014:GGP:2685048.2685096}.
It is a graph processing framework built on top of Apache Spark, enabling low-cost fault-tolerance.
The authors target graph processing by expressing graph-specific optimizations as distributed join optimizations and graph views' maintenance.
In GraphX, the property graph is reduced to a pair of collections.
This way, the authors are able to compose graphs with other collections in a distributed dataflow framework.
Operations such as adding additional vertex properties are then naturally expressed as joins against the collection of vertex properties.
Graph computations and comparisons are thus an exercise in analyzing and joining collections.
These tasks are routine in a broader graph analytics scope, but the graph-parallel abstraction is not the best way to satisfy them.

\end{itemize}

The last systems are in fact explicit storage systems for graph-specific data.
They all have different graph query APIs (some of which exist as standalone open source modules adopted by other projects) but with a common theme: graph traversals.
Rather than perform computation for each vertex of the graph, queries typically begin by referencing specific vertices by properties.
The query then results in a traversal beginning from the referenced vertices.

\begin{itemize}[leftmargin=*]

\item \textbf{Neo4j}.
NoSQL graph database implemented in \texttt{Scala} and \texttt{Java}.
There is a Community Edition licensed under the free GNU General Public License (GPL) v3.
Neo4j is optimized for highly-connected data.
It relies on methods of data access for graphs without considering data locality. 
Neo4j's graph processing consists of mostly random data access.
For large graphs which require out-of-memory processing, the major performance bottleneck becomes random secondary storage access.
Large-scale graphs need to be partitioned and distributed over several machines to reach scalable processing capabilities.
This is a centralized system that lacks the computational power of a distributed, parallel system.
The authors created a system which supports ACID transactions, high availability, with operations that modify data occurring within transactions to guarantee consistency.
As previously mentioned, it uses the query language Cypher.

\item \textbf{Titan}. It is an open-source project licensed under the Apache License 2.0.~\footnote{Access date 28 Oct, '16:~\url{https://github.com/thinkaurelius/titan/}}
Titan is a multi-machine cluster computing graph processing database~\footnote{Access date 28 Oct, '16:~\url{http://titan.thinkaurelius.com/}} which supports both ACID and eventual consistency.
It supports thousands of users performing graph traversals concurrently.
The project supports Cassandra and HBase as storage back-ends and integrates with big data analytics tools like Spark, Giraph and Hadoop.
The developers behind Titan are now working in a commercial solution, the DataStax Enterprise Graph\footnote{Access date 22 Dec, '16:~\url{http://www.datastax.com/products/datastax-enterprise-graph}}.

\item \textbf{Weaver}.
Presented in \textit{VLDB 2016}~\cite{DBLP:journals/corr/DubeyHES15}, it is an open-source project under a custom permissive license~\footnote{Access date 21 Nov, '16:~\url{https://github.com/dubey/weaver}}.
The authors describe it as a distributed graph database for efficient, transactional graph analytics.
They introduced the concept of refinable timestamps.
It is a mechanism to obtain a rough ordering of distributed operations if that is sufficient, but also fine-grained orderings when they become necessary.
Weaver supports transactions with strictly serialized ordering.
It is capable of distributing a graph across multiple shards while supporting concurrency.
Refinable timestamps allow for the existence of a multi-version graph: write operations use their timestamp as a mark for vertices and edges.
This allows for the existence of consistent versions of the graph so that long-running analysis queries to operate on a consistent version of the graph, as well as historical queries.
Weaver is written in \texttt{C++}, offering binding options for different languages.

\end{itemize}

\subsection{Storage Backends}\label{sec:sec:backends}
This subsection goes over the most relevant storage solutions one may find in the literature, as far as the author knows.
Many of the systems presented in Sec.~\ref{sec:sec:all_systems} provide high-level graph processing functionality and make use of these storage backends.
Some of the processing systems are even capable of interchanging these storage solutions.
It is important to remark that these solutions were not specifically built for graph-specific needs, although they were used as building blocks for such types of systems.
There are also storage backends with an explicit focus on graphs, but they are typically coupled with graph processing functionalities and semantics, which led to their inclusion in Sec.~\ref{sec:sec:all_systems} as variants of graph processing systems instead.
The backends are thus:

\begin{itemize}[leftmargin=*]

\item \textbf{BerkeleyDB}.
Debuted in \textit{ATEC 1999}~\cite{Olson:1999:BD:1268708.1268751} and maintained by Oracle.
It is a library that aims to support several useful building block technologies such as ACID transactions, non-blocking writes, concurrency and scalability, among others.
The authors aimed for developers to be able to incorporate it in a transparent way.
It has an API which supports a plethora of languages.

\item \textbf{HBase}. 
It is an open source, non-relational, distributed database modeled after Google's BigTable~\cite{Chang:2006:BDS:1267308.1267323} \textit{(2006)} and is written in \texttt{Java}~\cite{DBLP:books/daglib/0027893}.
Apache HBase is licensed under Apache License 2.0 and provides BigTable-like capabilities on top of Hadoop and HDFS~\cite{5496972}.
This project's goal is the hosting of very large tables -- billions of rows per millions of columns -- atop clusters of commodity hardware.
HBase doesn't support expressing queries as SQL scripts -- instead they are written in \texttt{Java}, sharing similarities with MapReduce.
It was designed to scale horizontally.
HBase works via a distinction of master/slave nodes in the cluster computing system.
It supports ACID-level semantics on a per-row basis.
This project relies on Apache ZooKeeper~\footnote{Access date 21 Oct, '16:~\url{https://zookeeper.apache.org/}} to perform inter-node communications.

\item \textbf{Cassandra}.
Originally presented in \textit{SIGOPS 2010}~\cite{Lakshman:2010:CDS:1773912.1773922}.
An open source\break project~\footnote{Access date 21 Oct, '16:~\url{https://github.com/apache/cassandra}.} under the Apache License 2.0, it is a distributed database management system designed by Facebook to handle large amounts of data across many commodity servers, providing high availability with no single point of failure.
Cassandra offers robust support for clusters spanning multiple data centers, with asynchronous (with no masters) replication allowing low latency operations for all clients.
Cassandra also places a high value on performance.
In 2012, University of Toronto researchers studying NoSQL systems discuss~\cite{Rabl:2012:SBD:2367502.2367512}:
\textit{
"In terms of scalability, there is a clear winner throughout our experiments.
Cassandra achieves the highest throughput for the maximum number of nodes in all experiments" although "this comes at the price of high write and read latencies."
}
Cassandra's data model is a partitioned row store with tunable consistency.
Rows are organized into tables; the first component of a table's primary key is the partition key; within a partition, rows are clustered by the remaining columns of the key. 
Other columns may be indexed separately from the primary key.
Tables may be created, dropped, and altered at run-time without blocking updates and queries.
Cassandra does not support joins or sub-queries, however.
Furthermore, the author notes that Cassandra requires the identification of some nodes \textit{as seed nodes}, to be used as points to concentrate inter-cluster-computing communication.
It integrates the Gossip protocol for communication between nodes.
Cassandra guarantees high availability by allowing a multiplicity of seed nodes in a cluster computing system.

\item \textbf{Alluxio}.
Formerly Tachyon~\cite{Li_2014_TRM_2670979.2670985} which was introduced in \textit{SOCC 2014}.
It is an open-source project~\footnote{Access date 21 Oct, '16:~\url{https://github.com/Alluxio/alluxio}} under the Apache License 2.0.
It is a distributed file system which enables reliable data sharing at memory speed across cluster computing frameworks.
The authors raise the issue that to support fault-tolerance, several systems (among them Cassandra and HBase) write at least one copy of data to non-volatile media. 
This is done in order to survive failures which span data centers.
However, disk limitations of write throughput coupled with in-memory computation frameworks amplify inter-job data sharing costs.
It is stated by the authors that to provide high throughput, fault-tolerance in storage systems must be achieved without replication.\\
\indent In the scope of this document, Alluxio also gains relevance because its authors delve into resource consumption from re-computations in the case of failures.
They devised a mathematical model and employed traces from Bing and Facebook to calculate the amount of resources spent on re-computation.
Overall, the authors conclude that lineage-based recovery could very well be the unique way to achieve memory-like throughput in cluster computing storage systems.

\end{itemize}

\section{Comparative System Assessment}\label{sec:study}

A listing of the major features and attributes of various systems in the literature was presented so far.
Due to the exhaustiveness of the list, a deep pair-wise comparative analysis of the systems is not practical.
However, based on qualitative aspects, the author remarks the prominence of three systems in particular: Apache Flink, Titan and Weaver.\\
\indent Among these three, upon considering their cases, it was decided to compare Apache Flink and Weaver.
The former has a vibrant community under active development with many built-in functionalities, while the later has shown promising performance results~\cite{DBLP:journals/corr/DubeyHES15} when compared to Titan.
While Titan has been incorporated into the DataStax range of products, the fact is that its open-source focus of development has dwindled.
Weaver's selling points were associated to its scalability as a distributed graph transaction system and its acclaimed usage to power the machine learning project RoboBrain (developed by the same institution and with a set of industrial and academic partners~\footnote{Access date 21 Oct, '16:~\url{http://robobrain.me/\#/people}.}).
While its adoption was small (when compared to Apache Flink), Weaver seemed to have a more recently-active community when compared to Titan, which was another point considered in its favor.

\subsection{Analyzed Systems, Datasets and Workloads}\label{sec:sec:datasets}

Apache Flink's implementation and testing in itself was straightforward.
Its community and countless usage examples were a solid basis to implement the invocation of the built-in graph algorithms included in its Gelly graph API.
It was first tested locally using \texttt{Java 8} to understand how it works, and then it was deployed in a Docker cluster computing system by adapting its built-in Docker tools which, despite some technical issues, allowed for smooth execution.\\
\indent However, the case with Weaver was much different.
Setting up Weaver was not trivial due to two reasons.\\
\indent First, to try out a local test of its API, it was found that its source code has dependencies which are not up-to-date with modern Linux distribution package versions.
As a consequence, an attempt was made to compile Weaver's \texttt{C++} code base from scratch.
However, this lead to the problem of missing package distributions.
Further effort was put into trying to compile the dependencies from source code, but the same problem was encountered with second-level package dependencies.\\
\indent At this point, effort was shifted to testing the Docker image provided by the authors of Weaver.
This image, contrary to what was expected, did not work out of the box.
The instructions given for this image were based on an older version of Docker, so an adaptation was made to start the Weaver image in the new version (instead of using a command-line call to launch each container, specifications were placed into a Dockerfile).
Weaver's Docker image was configured so that its elements (system coordinator, workers and timestampers) were all listening on the localhost, assuming an execution context where each running Docker container was directly using the host machine's localhost network.
To isolate the Weaver images from the host and use Docker's private network mechanisms, it was necessary to specify different addresses in each running container's Weaver configurations.
To edit Weaver's address and port listening values, it was necessary to look into its source code to understand where and how to provide the values.
The documentation and support groups of Weaver did not help in any way whatsoever in this regard.\\
\indent After successfully setting up a Weaver cluster computing system over Docker (and ensuring its nodes were communicating between themselves), further technical difficulties were found when trying to test its API.
Many of the API tests included in the Weaver image failed when running out of the box.
Some of the examples were built for a previous version of the Weaver API, but were still included in the image with no notice of deprecation.
As such, additional time was put into understanding if these failures stemmed from the Docker configuration changes that were made or if they were caused by problems inherent to the distributed Weaver.
Furthermore, the set of graph algorithms included in the Weaver distribution (dubbed \texttt{node\_programs}) made use of internal \texttt{C++} message-passing functionality.
Overall, the presented (and focused by the authors) \texttt{Python} API's actual functionalities were not enough to implement and test a simple label propagation algorithm, at least not without investing further time in writing a \texttt{C++} \texttt{node\_program} and recompiling Weaver.
This was very detrimental to one of the advantages of using the Docker image -- shipping with binaries and ready to run.\\
\indent As a consequence of the time constraints incurred by the lack of maintenance when exploring Weaver, it was decided to focus on Apache Flink 1.1.3. 
\begin{table}
    \centering
	\begin{tabular}{ |c|c|c|c| }
		\hline 
		$D$ & $|$V$|$ & $|$E$|$ & \#$C$\\ \hline 
		\multicolumn{1}{|l}{com-dblp.ungraph.txt} & \multicolumn{1}{|r}{317,080} & \multicolumn{1}{|r}{1,049,866} & \multicolumn{1}{|r|}{13,477}\\ \hline
		\multicolumn{1}{|l}{com-orkut.ungraph.txt} & \multicolumn{1}{|r}{3,072,441} & \multicolumn{1}{|r}{117,185,083} & \multicolumn{1}{|r|}{6,288,363}\\ \hline
		\multicolumn{1}{|l}{com-friendster.ungraph.txt} & \multicolumn{1}{|r}{65,608,366} & \multicolumn{1}{|r}{1,806,067,135} & \multicolumn{1}{|r|}{957,154}\\ \hline
	\end{tabular}
	\caption{A heterogeneous set of undirected graphs was considered for testing purposes.
They are part of the Stanford Network Analysis Platform (SNAP) datasets with ground-truth communities.
The first line was the smallest dataset in SNAP, the DBLP network. With progressively increasing sizes in both node count $|$V$|$ and edge count $|$E$|$, DBLP is followed by the Orkut and Friendster community networks. The last column for each dataset represents the number of communities identified by~\cite{DBLP:journals/corr/abs-1205-6233}.}\label{table:datasets}
\end{table}
The datasets chosen for testing are summarized in Table~\ref{table:datasets}.
Only the largest of these datasets, the Friendster community, was chosen for benchmarking.
This was due to the sub-minute execution times obtained for $W=1$ setups on the DBLP and Orkut datasets.
Despite not being used, they were presented in Table~\ref{table:datasets} to aid in understanding the relativity of the graph sizes.
While establishing ground-truth was not a mission-critical objective, care was taken to ensure such data was available~\footnote{Access date 15 Dec, '16:~\url{https://snap.stanford.edu/data/\#communities}} in case any particular validation was needed.
These datasets were featured in a study~\cite{DBLP:journals/corr/abs-1205-6233} which researched how different metrics (such as clustering coefficient) had an impact on detected communities.
An attempt to exactly reproduce the results of~\cite{DBLP:journals/corr/abs-1205-6233} is not made, as their detection methods use various algorithms as sub-steps for discarding and validating calculated communities.
The focus here was cast on analyzing the performance in a controlled distributed system scenario.\\
\indent This section thus covers and discusses key-points in the experimental comparison of two built-in community detection algorithms in Apache Flink.
The algorithms targeted were: \textit{1)} score-based label propagation~\cite{PhysRevE.76.036106} (it shall be referred to as LAx) and \textit{2)} a variant that does not use scores~\cite{PhysRevE.79.066107} (it shall be referred to as LA).

\subsection{Implementation and Experimental Setup}\label{sec:sec:experimenting}

The two tested algorithms~\footnote{Apache Flink 1.1.3 API reference:\\
LA: \texttt{org.apache.flink.graph.library.LabelPropagation}\\
LAx: \texttt{org.apache.flink.graph.library.CommunityDetection}} were configured both to run until 10 iterations were executed or if no changes occurred between iterations.
The platform used for all testing scenarios was a Docker-based cluster computing system with a varying number of workers $W$, with each system execution for a given dataset $D$ being performed $R$ times.
For this study, experiments had a fixed $R=10$ and a computing cluster size varying from $W=1$ to $W=10$.
Essentially, the presented execution times show the average execution time (which is bounded by the slowest worker) for an increasing worker count $W$ and a constant number of repetitions $R$.
Presented error bars are based on the standard deviation obtained for each of the previously mentioned parameters.
These figures were generated with the \texttt{gnuplot 5.0} utility.
Changes were implemented with respect to copying a label propagation \texttt{Java} client program into the images.\\
\indent The Docker image provided in the Apache Flink distribution underwent minimal modification.
This program was implemented during this project and it is through it that the skeleton of the dataflow program in Flink is defined.\\
\indent The computational infrastructure used was an SMP with 256 GB RAM and 8 Intel(R) Xeon(R) CPU E7- 4830 @ 2.13GHz with eight cores each.
Care was taken to ensure, whenever possible, that the system's cluster configuration and container nodes responsible for processing had equal amounts of working memory (4 GB for worker nodes, 2 GB for cluster computing coordinator-type nodes) available to them.
This served the purpose of ensuring a uniform testing scenario with minimal variation.
It is also relevant to mention that all presented data was obtained during periods where overall infrastructure usage was homogeneous - again, this was to dampen yet another possible source of result skew.

\subsection{Discussion}\label{sec:sec:discussion}

\begin{figure}
    \centering
    \includegraphics[width=0.75\columnwidth]{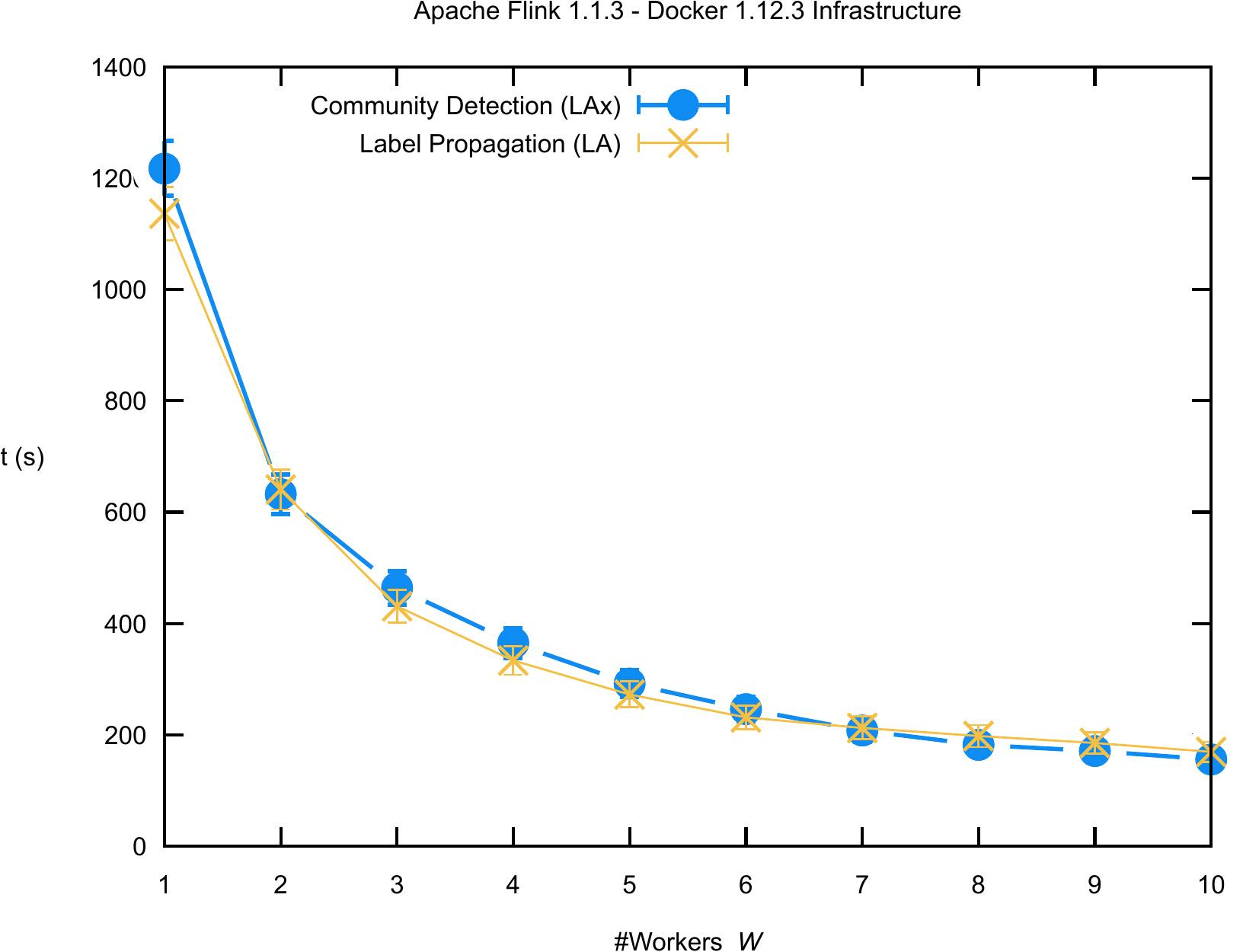}
    \caption{Execution times for the SNAP Friendster community dataset.}
    \label{fig:friendster-time}
\end{figure}

Algorithm execution times are presented in Fig.~\ref{fig:friendster-time}.
As one would expect, initial increments in worker count $W$ produce close to linear performance improvements.
However, after a certain point, diminishing gains are incurred.
This is due to the relation between dataset size and communication overheads.
For an even bigger dataset, one would intuitively expect that the diminishing returns would be attenuated.
Intuitively, one may expect these datasets to manifest behaviors which lead to different communication overheads in distributed system scenarios.
Despite using an underlying virtualization to simulate a cluster computing system over a single machine, it would be expected to obtain less than linear speedup for the same dataset as executions are performed with more workers.
Recall Amdahl's Law~\cite{Amdahl:1967:VSP:1465482.1465560}, which states that the theoretical speedup $\phi(W)$ for $W$ workers and a fraction $P$ of the execution that can be made parallel may be obtained as such:

\begin{equation}
\phi(W) = \frac{1}{(1 - P) + \frac{P}{W}}
\end{equation}

\begin{figure}
    \centering
    \includegraphics[width=0.75\columnwidth]{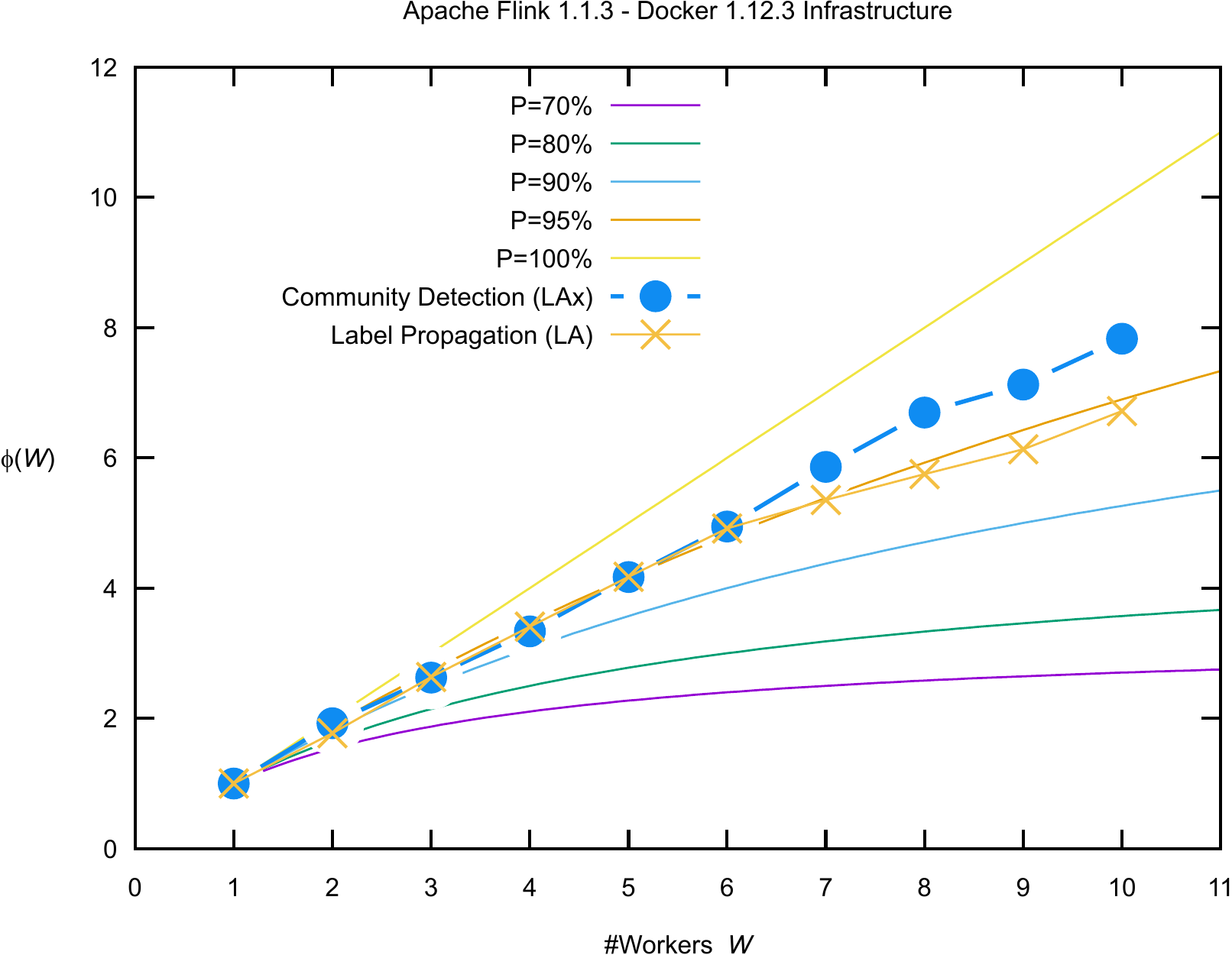}
    \caption{Speedup results $\phi$ with respect to the temporal data of the SNAP Friendster community dataset.}
    \label{fig:friendster-speedup}
\end{figure}

Using this law, different configurations of Amdahl's Law are plotted in Fig.~\ref{fig:friendster-speedup} with different values for the $P$ fraction.
In the figure, there are plots of the theoretical speedup obtained by Amdahl's law for different assumed values of the percentage $P$ of execution that benefits from parallelism.
Both algorithms are very close to a theoretical speedup curve associated to 95\% of the execution benefiting from parallelism.
It is important to note that the Docker-based infrastructure will have minimized worker communication latencies.
There is the possibility that achieved speedup values would be penalized with greater communication overheads in a real physical cluster computing infrastructure evaluation.
This approach was used as a basis for inferring the bounds of the empirical fraction $P$ of execution that can be made parallel, with respect to the two evaluated algorithms.
While moving from one to two workers approximately doubles performance, adding further workers did not yield a linear speedup.
Figure~\ref{fig:friendster-speedup} thus shows speedup values associated with the times of Fig~\ref{fig:friendster-time} plus theoretical plots of Amdahl's Law for varying parameter $P$.
It is important to notice that the communication overhead in this setup is not the worst-case scenario one could expect, as cluster computing system communication is bound to be faster inside an SMP than in an actual physical cluster.\\
\indent Overall, the showcased results indicate that these two variants of community-detecting algorithms have an execution overhead of around 5\%.
This is in part due to communication costs (though they are reduced in a cluster computing system) and the synchronization mechanisms in Apache Flink, which are bound to be efficient in order to reach such a low number.
On average, worker node memory usage was around 4 GB for both algorithm scenarios.

\section{Conclusion}\label{sec:conclusion}

The experimental work conducted in this study aimed at gaining preliminary insights on the most desirable platforms for further studies.
Distributed graph processing has a rich set of applications, and the number of systems that have been built for it allow users to express computations in different ways.\\
\indent As an example, one may express graph computation in Apache Flink in different ways, such as the vertex-centric approach. 
It is over that approach that the built-in label propagation algorithm is expressed.\\
\indent While this work allowed for gaining experience on the configuration of a cluster computing system for graph processing, a considerable amount of time went into obtaining a uniform testing basis -- Docker was employed as a means to this end.
While there was an initial overhead due to learning and configuring Docker for Apache Flink and Weaver, the experience gained will prove useful for evaluating additional distributed graph processing systems and fine-tuning controlled execution scenarios. 

\section{Future Work}\label{sec:future}

Graph data is present in many domains -- reducing computational costs and achieving greater performance on graphs will thus allow us to reach greater energy efficiency.
The study conducted in this work highlighted different aspects of existing graph processing platforms and elected Apache Flink as a strong candidate for further research activities.\\
\indent Studies have been done on establishing a set of semi-metrics upon which other graph properties depend~\cite{kalavri2016shortest}.
Along this line, it would be relevant to explore how statistics and machine learning can guarantee that results and metrics can be kept up-to-date with minimal overheads and performance impacts.
Researching these techniques and the use of graph-specific databases incorporated into a platform such as Apache Flink makes for a logical vector for future research.

\bibliographystyle{splncs03}
\bibliography{paper}

\begin{thebibliography}{10}
\providecommand{\url}[1]{\texttt{#1}}
\providecommand{\urlprefix}{URL }

\bibitem{Amdahl:1967:VSP:1465482.1465560}
Amdahl, G.M.: Validity of the single processor approach to achieving large
  scale computing capabilities. In: Proceedings of the April 18-20, 1967,
  Spring Joint Computer Conference. pp. 483--485. AFIPS '67 (Spring), ACM, New
  York, NY, USA (1967)

\bibitem{Blondel08fastunfolding}
Blondel, V.D., loup Guillaume, J., Lambiotte, R., Lefebvre, E.: Fast unfolding
  of communities in large networks (2008)

\bibitem{Boldi:2004:WFI:988672.988752}
Boldi, P., Vigna, S.: The webgraph framework i: Compression techniques. In:
  Proceedings of the 13th International Conference on World Wide Web. pp.
  595--602. WWW '04, ACM, New York, NY, USA (2004)

\bibitem{Boldi:2011:LLP:1963405.1963488}
Boldi, P., Rosa, M., Santini, M., Vigna, S.: Layered label propagation: A
  multiresolution coordinate-free ordering for compressing social networks. In:
  Proceedings of the 20th International Conference on World Wide Web. pp.
  587--596. WWW '11, ACM, New York, NY, USA (2011)

\bibitem{Boldi03thewebgraph}
Boldi, P., Vigna, S.: The webgraph framework i: Compression techniques. In: In
  Proc. of the Thirteenth International World Wide Web Conference. pp.
  595--601. ACM Press (2003)

\bibitem{ACombinatorialProblem1946}
de~Bruijn, N.G.: {A Combinatorial Problem}. Koninklijke Nederlandsche Akademie
  Van Wetenschappen  49(6),  758--764 (Jun 1946)

\bibitem{Chang:2006:BDS:1267308.1267323}
Chang, F., Dean, J., Ghemawat, S., Hsieh, W.C., Wallach, D.A., Burrows, M.,
  Chandra, T., Fikes, A., Gruber, R.E.: Bigtable: A distributed storage system
  for structured data. In: Proceedings of the 7th USENIX Symposium on Operating
  Systems Design and Implementation - Volume 7. pp. 15--15. OSDI '06, USENIX
  Association, Berkeley, CA, USA (2006)

\bibitem{Ching:2015:OTE:2824032.2824077}
Ching, A., Edunov, S., Kabiljo, M., Logothetis, D., Muthukrishnan, S.: One
  trillion edges: Graph processing at facebook-scale. Proc. VLDB Endow.  8(12),
   1804--1815 (Aug 2015)

\bibitem{dimiduk2012hbase}
Dimiduk, N., Khurana, A., Ryan, M.: HBase in Action. Running Series, Manning
  (2012)

\bibitem{DBLP:journals/corr/DubeyHES15}
Dubey, A., Hill, G.D., Escriva, R., Sirer, E.G.: Weaver: {A} high-performance,
  transactional graph store based on refinable timestamps. CoRR  abs/1509.08443
  (2015)

\bibitem{Fortunato:2007p5549}
Fortunato, S., Barthelemy, M.: Resolution limit in community detection.
  Proceedings of the National Academy of Sciences  (Jan 2007)

\bibitem{DBLP:books/daglib/0027893}
George, L.: HBase - The Definitive Guide: Random Access to Your Planet-Size
  Data. O'Reilly (2011)

\bibitem{Gonzalez:2014:GGP:2685048.2685096}
Gonzalez, J.E., Xin, R.S., Dave, A., Crankshaw, D., Franklin, M.J., Stoica, I.:
  Graphx: Graph processing in a distributed dataflow framework. In: Proceedings
  of the 11th USENIX Conference on Operating Systems Design and Implementation.
  pp. 599--613. OSDI'14, USENIX Association, Berkeley, CA, USA (2014)

\bibitem{Han:2014:ECP:2732977.2732980}
Han, M., Daudjee, K., Ammar, K., \"{O}zsu, M.T., Wang, X., Jin, T.: An
  experimental comparison of pregel-like graph processing systems. Proc. VLDB
  Endow.  7(12),  1047--1058 (Aug 2014)

\bibitem{jones2004introduction}
Jones, N., Pevzner, P.: An Introduction to Bioinformatics Algorithms. A
  Bradford book, London (2004)

\bibitem{kalavri2016shortest}
Kalavri, V., Simas, T., Logothetis, D.: The shortest path is not always a
  straight line: leveraging semi-metricity in graph analysis. Proceedings of
  the VLDB Endowment  9(9),  672--683 (2016)

\bibitem{10.2307/2033241}
Kruskal, J.B.: On the shortest spanning subtree of a graph and the traveling
  salesman problem. Proceedings of the American Mathematical Society  7(1),
  48--50 (1956)

\bibitem{Lakshman:2010:CDS:1773912.1773922}
Lakshman, A., Malik, P.: Cassandra: A decentralized structured storage system.
  SIGOPS Oper. Syst. Rev.  44(2),  35--40 (Apr 2010)

\bibitem{leskovec2014mining}
Leskovec, J., Rajaraman, A., Ullman, J.: Mining of Massive Datasets. Cambridge
  University Press (2014), \url{https://books.google.pt/books?id=1l-WoAEACAAJ}

\bibitem{PhysRevE.79.066107}
Leung, I.X.Y., Hui, P., Li\`o, P., Crowcroft, J.: Towards real-time community
  detection in large networks. Phys. Rev. E  79,  066107 (Jun 2009)

\bibitem{Li_2014_TRM_2670979.2670985}
Li, H., Ghodsi, A., Zaharia, M., Shenker, S., Stoica, I.: Tachyon: Reliable,
  memory speed storage for cluster computing frameworks. In: Proceedings of the
  ACM Symposium on Cloud Computing. pp. 6:1--6:15. SOCC '14, ACM, New York, NY,
  USA (2014)

\bibitem{Li01022010}
Li, R., Zhu, H., Ruan, J., Qian, W., Fang, X., Shi, Z., Li, Y., Li, S., Shan,
  G., Kristiansen, K., Li, S., Yang, H., Wang, J., Wang, J.: De novo assembly
  of human genomes with massively parallel short read sequencing. Genome
  Research  20(2),  265--272 (2010)

\bibitem{Low_graphlab:a}
Low, Y., Bickson, D., Gonzalez, J., Guestrin, C., Kyrola, A., Hellerstein,
  J.M.: Graphlab: A new framework for parallel machine learning

\bibitem{Low:2012:DGF:2212351.2212354}
Low, Y., Bickson, D., Gonzalez, J., Guestrin, C., Kyrola, A., Hellerstein,
  J.M.: Distributed graphlab: A framework for machine learning and data mining
  in the cloud. Proc. VLDB Endow.  5(8),  716--727 (Apr 2012)

\bibitem{Malewicz:2010:PSL:1807167.1807184}
Malewicz, G., Austern, M.H., Bik, A.J., Dehnert, J.C., Horn, I., Leiser, N.,
  Czajkowski, G.: Pregel: A system for large-scale graph processing. In:
  Proceedings of the 2010 ACM SIGMOD International Conference on Management of
  Data. pp. 135--146. SIGMOD '10, ACM, New York, NY, USA (2010)

\bibitem{Malicevic:2014:SGP:2592784.2592789}
Malicevic, J., Roy, A., Zwaenepoel, W.: Scale-up graph processing in the cloud:
  Challenges and solutions. In: Proceedings of the Fourth International
  Workshop on Cloud Data and Platforms. pp. 5:1--5:6. CloudDP '14, ACM, New
  York, NY, USA (2014)

\bibitem{Murphy:2012:MLP:2380985}
Murphy, K.P.: Machine Learning: A Probabilistic Perspective. The MIT Press
  (2012)

\bibitem{nathan1967inversion}
Nathan, A., Even, R.K.: The inversion of sparse matrices by a strategy derived
  from their graphs. The Computer Journal  10(2),  190--194 (1967)

\bibitem{Olson:1999:BD:1268708.1268751}
Olson, M.A., Bostic, K., Seltzer, M.: Berkeley db. In: Proceedings of the
  Annual Conference on USENIX Annual Technical Conference. pp. 43--43. ATEC
  '99, USENIX Association, Berkeley, CA, USA (1999)

\bibitem{BLTJ:BLTJ1515}
Prim, R.C.: Shortest connection networks and some generalizations. Bell System
  Technical Journal  36(6),  1389--1401 (1957)

\bibitem{Rabl:2012:SBD:2367502.2367512}
Rabl, T., G\'{o}mez-Villamor, S., Sadoghi, M., Munt{\'e}s-Mulero, V., Jacobsen,
  H.A., Mankovskii, S.: Solving big data challenges for enterprise application
  performance management. Proc. VLDB Endow.  5(12),  1724--1735 (Aug 2012)

\bibitem{PhysRevE.76.036106}
Raghavan, U.N., Albert, R., Kumara, S.: Near linear time algorithm to detect
  community structures in large-scale networks. Phys. Rev. E  76,  036106 (Sep
  2007)

\bibitem{PhysRevE.81.046114}
Ronhovde, P., Nussinov, Z.: Local resolution-limit-free potts model for
  community detection. Phys. Rev. E  81,  046114 (Apr 2010),
  \url{http://link.aps.org/doi/10.1103/PhysRevE.81.046114}

\bibitem{Roy:2015:CSG:2815400.2815408}
Roy, A., Bindschaedler, L., Malicevic, J., Zwaenepoel, W.: Chaos: Scale-out
  graph processing from secondary storage. In: Proceedings of the 25th
  Symposium on Operating Systems Principles. pp. 410--424. SOSP '15, ACM, New
  York, NY, USA (2015)

\bibitem{Roy:2013:XEG:2517349.2522740}
Roy, A., Mihailovic, I., Zwaenepoel, W.: X-stream: Edge-centric graph
  processing using streaming partitions. In: Proceedings of the Twenty-Fourth
  ACM Symposium on Operating Systems Principles. pp. 472--488. SOSP '13, ACM,
  New York, NY, USA (2013)

\bibitem{5496972}
Shvachko, K., Kuang, H., Radia, S., Chansler, R.: The hadoop distributed file
  system. In: 2010 IEEE 26th Symposium on Mass Storage Systems and Technologies
  (MSST). pp. 1--10 (May 2010)

\bibitem{Tian:2013:TLV:2732232.2732238}
Tian, Y., Balmin, A., Corsten, S.A., Tatikonda, S., McPherson, J.: From "think
  like a vertex" to "think like a graph". Proc. VLDB Endow.  7(3),  193--204
  (Nov 2013)

\bibitem{Waltman2013}
Waltman, L., van Eck, N.J.: A smart local moving algorithm for large-scale
  modularity-based community detection. The European Physical Journal B
  86(11),  1--14 (2013)

\bibitem{DBLP:journals/corr/abs-1205-6233}
Yang, J., Leskovec, J.: Defining and evaluating network communities based on
  ground-truth. CoRR  abs/1205.6233 (2012)

\bibitem{Zaharia:2010:SCC:1863103.1863113}
Zaharia, M., Chowdhury, M., Franklin, M.J., Shenker, S., Stoica, I.: Spark:
  Cluster computing with working sets. In: Proceedings of the 2Nd USENIX
  Conference on Hot Topics in Cloud Computing. pp. 10--10. HotCloud'10, USENIX
  Association, Berkeley, CA, USA (2010)

\bibitem{18349386}
Zerbino, D., Birney, E.: Velvet: algorithms for de novo short read assembly
  using de bruijn graphs. Genome Res  18,  821 (2008)

\end{thebibliography}

\end{document}